\documentclass[sigconf]{acmart}

\usepackage{pdfpages}
\usepackage{caption,subcaption}
\usepackage{graphicx}
\usepackage{array} 
\usepackage{capt-of}
\usepackage{booktabs}
\usepackage{varwidth}

\AtBeginDocument{%
  }

\setcopyright{none}
\acmDOI{}
\acmISBN{}
\acmConference[LIMITS '25]{11th Workshop on Computing within Limits}{June 26--27, 2025}{Online}

\begin{document}

\title{Exploring the Viability of the Updated World3 Model for Examining the Impact of Computing on Planetary Boundaries}

\author{Nara Guliyeva}
\email{nargiz.guliyeva@mail.utoronto.ca}
\affiliation{%
  \institution{University of Toronto}
  \city{Toronto}
  \state{Ontario}
  \country{Canada}
}

\author{Eshta Bhardwaj}
\email{eshta.bhardwaj@mail.utoronto.ca}
\affiliation{%
  \institution{University of Toronto}
  \city{Toronto}
  \state{Ontario}
  \country{Canada}
}

\author{Christoph Becker}
\email{christoph.becker@utoronto.ca}
\affiliation{%
  \institution{University of Toronto}
  \city{Toronto}
  \state{Ontario}
  \country{Canada}
}

\renewcommand{\shortauthors}{Guliyeva et al.}

\begin{abstract}
  The influential \textit{Limits to Growth} report introduced a system dynamics-based model to demonstrate global dynamics of the world’s population, industry, natural resources, agriculture, and pollution between 1900-2100. In current times, the rapidly expanding trajectory of data center development, much of it linked to AI, uses increasing amounts of natural resources. The extraordinary amount of resources claimed warrants the question of how computing trajectories contribute to exceeding planetary boundaries. Based on the general robustness of the World3-03 model and its influence in serving as a foundation for current climate frameworks, we explore whether the model is a viable method to quantitatively simulate the impact of data centers on limits to growth. Our paper explores whether the World3-03 model is a feasible method for reflecting on these dynamics by adding new variables to the model in order to simulate a new AI-augmented scenario. We find that through our addition of AI-related variables (such as increasing data center development) impacting pollution in the World3-03 model, we can observe the expected changes to dynamics, demonstrating the viability of the World3-03 model for examining AI’s impact on planetary boundaries. We detail  future research opportunities for using the World3-03 model to explore the relationships between increasing resource-intensive computing and the resulting impacts to the environment in a quantitative way given its feasibility. 

\end{abstract}



\maketitle

\section{Introduction}

The 1972 \textit{Limits to Growth} (LtG) report \cite{meadows_limits_1972} used a global model, called World3, to show how economic growth would be limited by both increasing population and usage of natural resources and represented this through 12 potential future scenarios. Their World3 model was based on system dynamics, an approach to systems thinking which uses ``measurable variables [to] represent structural properties of the world... and organizes these variables with quantitative causal relationships that allow it to predict overall behavior as the emergent property'' \cite[p.~20]{christoph_becker_insolvent_2023}. The model thus provided a method for exploring the dynamics between the ``expanding global population and materials economy'' and ``the [Earth’s] limited carrying capacity'' \cite{donella_meadows_synopsis_2004}. Given that the model and report were first introduced in 1972 \cite{meadows_limits_1972}, with updates in 1992 \cite{meadows_beyond_1992} and 2004 \cite{meadows_limits_2004}, a consideration of then-emerging concepts of `AI' as a technological development with global ecological relevance was expectedly not considered. 

In current times, the high computational training requirements of so-called `AI' drive substantial energy consumption \cite{crawford_generative_2024, luccioni_power_2024,sayegh_billion-dollar_2024,sevilla_can_2024}, manufacturing demands \cite{luccioni_environmental_2024, muldoon_poverty_2023}, and investment needs (e.g., \cite{herrman_ai_2024,nestor_maslej_ai_2024}, which could significantly impact resource depletion \cite{bhardwaj_limits_2025, luccioni_environmental_2024, Brevini_2021} and pollution levels globally \cite{han_unpaid_2024}. The trajectory and broader implications of the recent surge in AI are presently widely researched with several studies showcasing that the global proliferation of AI models intensifies environmental pressures \cite{bhardwaj_limits_2025,guidi_environmental_2024,luccioni_environmental_2024,varoquaux_hype_2025,Luccioni_Strubell_Crawford_2025}, thereby necessitating further exploration of the extent of this impact.

Accordingly, this paper explores whether the effect of vastly scaling AI on patterns of world development, particularly environmental outcomes, can be quantified using the World3-03 model. Feedback loops resulting from AI advancements that rely heavily on energy-intensive data centers \cite{dayarathna_data_2015,eric_masanet_how_2020,guidi_environmental_2024,olivo_internet_2024}, land use \cite{muldoon_poverty_2023}, rare and finite raw materials for hardware \cite{luccioni_environmental_2024}, vast datasets for training \cite{sevilla_compute_2022}, and substantial financial investments (e.g., \cite{keach_hagey_sam_2024,kimball_microsoft_2024,nestor_maslej_ai_2024,ondrej_burkacky_semiconductor_2022,sayegh_billion-dollar_2024,wong_silicon_2024}) could accelerate (or mitigate) resource depletion and fundamentally reshape sustainability dynamics. 

We examine the following research question: 

\begin{enumerate}
    \item Is the \textit{Limits to Growth} World3-03 model a viable method for exploring the impact of computing on planetary boundaries today? 
\end{enumerate}

We answer this question by specifically exploring whether and how AI development trajectories can be incorporated into the World3-03 model, which is key to anticipating its long-term consequences on resource availability and stability. We focus here specifically on direct lifecycle impacts resulting from the materials and operations of data centers. We do this by adding new variables to the World3-03 model in order to simulate impacts on pollution as a result of the accelerating development of data centers. Given the rise in data center development causing increased pollution presently and estimated to skyrocket in the future \cite{han_unpaid_2024}, our hypothesis is that by adding variables and setting parameters that increase pollution, we are able to simulate consequences to the relevant stocks in the World3-03 model. Thus we quantify assumptions regarding data center-caused pollution within the World3-03 model to simulate the expected consequences. Our findings confirm our hypothesis and illustrate a method for adjusting the World3-03 model to reflect ecological impact of computing.

We focus on variables that represent data center development to the World3-03 model for a few reasons. Our goal is to explore whether the World3-03 model can quantitatively simulate the impact of computing on limits to growth. Presently, computing's impacts are discussed and debated most widely pertaining to AI. 

`AI' is most commonly a marketing term rather than a technical term but currently most often refers to machine learning and large language models, often in the form of chatbots and generative models producing audiovisual and textual materials, including code. While references to AI often encompass things as diverse as the AI industry, AI models, systems, and technologies including generative AI, and the algorithms underlying AI systems, in this paper we focus on how the sector manifests as physical infrastructure, including the accelerating development and operations of hyperscale data centers. Our scope is to explore how data center development may impact pollution (a limit to growth recognized as a planetary boundary today) on a global scale. This is therefore one method by which we can explore how AI's impact can be added to the World3-03 model. 

Through reviving the World3-03 model, the paper encourages further examination of the links between computing and its environmental footprint while identifying new research opportunities. By doing so, it provides the groundwork for future, comprehensive system dynamics perspectives on whether and how accelerating AI and computing broadly push humanity closer to, or further from, planetary boundaries.

\section{Background}

In this section, we first provide an overview of the findings of the \textit{Limits to Growth} report and the development of the World3-03 model. Next, we briefly touch on the critiques surrounding the report’s results, noting that some of these critiques were later revealed to rest on shaky grounds. Finally, we discuss the recent updates to the report and model and its continued general validity recognized throughout the recalibrations. 

\subsection{Limits to Growth and the World3-03 Model}

Pioneered by the \textit{Limits to Growth} study in 1972 \cite{meadows_limits_1972} and followed by two updates in 1992 \cite{meadows_beyond_1992} and 2004 \cite{meadows_limits_2004}, the World3 model is an application of a system dynamics-based approach to explore the interplay of population, industry, resources, agriculture, and pollution over long periods under resource constraints. The model was developed based on Forrester's exploration of the possibility to build a high-level simulation model of natural resource depletion, pollution crisis, crowding, and food shortage \cite{Forrester_1971}.

The equations and technical rationale behind the World3 model are presented in a separate volume, \textit{Dynamics of Growth in a Finite World} \cite{meadows_beyond_1993}. The original (1972) study explores 12 scenarios, which have been described in the Appendix A, Table 4. The debate was coined around 4 key scenarios: Business-as-Usual (BAU) -- Standard Run, Business-as-Usual 2 (BAU2) -- High Resources, Comprehensive Technology (CT) -- Tech Fix, and Stabilized World (SW), each illustrating different pathways for the future \cite{meadows_limits_1972}, explicated further in Table \ref{tab:tab1}. The 1972 LtG  uses the term ``run'' to describe various model simulations while the 2004 study uses the term ``scenario'' to refer to the same type of model-based exploration. This paper uses the term ``scenario'' for consistency.

\begin{table}
    \centering
\caption{Assumptions of World3-03 key scenarios, adapted from \cite{herrington_update_2021}.}
\label{tab:tab1}
    \begin{tabular}{|>{\raggedright\arraybackslash}p{0.25\linewidth}|>{\raggedright\arraybackslash}p{0.25\linewidth}|>{\raggedright\arraybackslash}p{0.42\linewidth}|} \hline 
         Scenario&  Description&  Cause\\ \hline 
         Business-as-Usual (BAU) – Standard Run&  No assumptions added to historic averages.&  Food production, industrial output, and population expand exponentially until the sharp depletion of resources eventually curtails industrial growth. Collapse is eventually caused by natural resource depletion.\\ \hline 
         Business-as-Usual 2 (BAU2) – High Resources&  Double the natural resources of BAU.&  Rapid population growth raises death rates and lowers food production, leaving resources severely depleted despite an initial doubling. Collapse is caused by pollution.\\ \hline 
 Comprehensive Technology (CT) – Tech Fix & BAU2, plus  exceptionally high technological development and adoption rates.&The technology delays the onset of limits significantly. Rising costs for technology eventually cause declines such as in industrial output, but no collapse.\\ \hline 
 Stabilized World (SW)& CT + changes in societal values and priorities.&The result of sustainability related policies would be population stabilization, as  human welfare is  on a high level. In this scenario, collapse is not expected. \\ \hline
    \end{tabular}

\end{table}

These scenarios emphasize a central message: the planet has a finite carrying capacity (i.e., a maximum population and consumption load that the ecosystem can support indefinitely without irreversible degradation) and thus requires a transition from growth to long-term sustainability.

In their 30-year update, Meadows et al. further reinforced that even by their 1992 update, humanity had already entered overshoot, exceeding the Earth’s carrying capacity, with climate change emerging as a major driver of systemic collapse \cite{meadows_limits_2004}. Unlike the 1972 edition, their 2004 book explored the results as a choice, highlighting ``visioning, networking, truth-telling, learning, and loving'' as social tools equal in importance to technological efficiency \cite{meadows_limits_2004}. Furthermore, the book conveys the need for continual model revision, a message that is directly relevant to our effort to incorporate current computing impacts into World3-03 \cite{meadows_limits_2004}. 

The 2004 book further described 10 scenarios that are similar to those outlined in the 1972 and 1992 versions \cite{meadows_limits_2004}. This version of the study presented the revised model, World3-03, which included new variables such as the human ecological footprint and human welfare. Furthermore, the updated study emphasized the growing importance of technology and its effect on the model. It focused less on exact prediction and more on pattern recognition and underscored that technological advancements alone are not sufficient to prevent collapse \cite{meadows_limits_2004}. While originally published in 1972, the \textit{Limits to Growth} report remains a relevant and vital influence in this area of study given its usage as a foundation for now influential frameworks such as the planetary boundaries \cite{rockstrom_planetary_2009, steffen_planetary_2015, Rockström_Kotzé_Milutinović_Biermann_Brovkin_Donges_Ebbesson_French_Gupta_Kim_2024} and doughnut economics \cite{raworth_doughnut_2017}.

\subsection{Criticism about the Limits to Growth Report}

Several critiques of the LtG surfaced soon after its 1972 release, and those were widely echoed by the general public in popular outlets, such as \cite{Passell_Ross_1972}. The LtG was scrutinized by economists and policy analysts eager to test the robustness of its assumptions and outputs. The critique ranged from methodological concerns to economic arguments that feedback and technological innovation were insufficiently represented. However, subsequent empirical and methodological work has largely overturned those criticisms. 

A comprehensive critique of the LtG challenges its methodology, data inputs, and the lack of context \cite{cole_models_1973}. Two overarching claims were made: 1) LtG has elected pessimistic parameter values; and 2) the model was itself flawed because small shifts in a few coefficients could swing the results from growth to collapse. Despite some valid points such as that testing models reveal errors in the results, the technical criticism was based on a misunderstanding of feedback-dominated models. For example, the claim that moving the initial year to 1850 would shift model’s collapse forward by twenty years ignores the need to recalibrate the entire parameter set to known historical data \cite{bardi_limits_2011}. It was also argued that the model could not be validated scientifically because its high-level aggregates were abstractions with no direct empirical outputs \cite{cole_models_1973}. However, the empirical results decades later have shown an alignment between several World3 scenarios and observed trends, suggesting that the model’s structural insights outweigh its parametric uncertainties \cite{herrington_update_2021,turner_comparison_2008}.

Another influential critique was made by Nordhaus who outlined his critique in three overlapping strands that have since framed much of the cross-disciplinary debate: 1) ad personam/epistemic posture (i.e., long-range simulation is ``hybris''), 2) unsubstantiated statements of disbelief (i.e., the values for equations are without any counter evidence); 3) quantifiable/econometric critique (i.e., due to the lack of econometric calibration, results are ``numerology'') \cite{nordhaus_world_1973}. Nordhaus' claims were rejected by Forrester \cite{forrester_debate_1974} with the counter arguments that Nordhaus had linearised inherently non-linear relationships and compared incomparable units when appealing to historical data. It was also highlighted that the purpose of LtG was not point forecasting but ``structured exploration'' of policy-sensitive feedback \cite{forrester_debate_1974}. After the 1992 update, Nordhaus made another critique that the LtG relies on a rigid system dynamics model that lacks empirical grounding and inadequately incorporates economic feedback and technological innovation \cite{Nordhaus_1992}. It is worth noting that Nordhaus' arguments rest on economic ideologies that lead to deeply problematic arguments \cite{gardiner_perfect_2014}, including the suggestion that the `optimal' temperature increase of the planet occurs at an average of 3.5 degrees Celsius -- a level typically described as `catastrophic' by climate scientists \cite{stern_economics_2022,ketcham_when_2023}. 

More recent criticism has stated that claims of systemic collapse are premature and should not be based solely on simulation outcomes and that the World3-03 model is overly abstract and simplified with uncertainties in key variables that are unreliable \cite{castro_arguments_2012}.

Bardi provides an analysis of the key conclusions of the LtG and its updates, arguing that despite criticism, the LtG's findings are valid \cite{bardi_limits_2011}. Bardi points out that academic criticism of the model stems from a misunderstanding of system dynamics itself and failing to grasp the model’s essence. Furthermore, the author states that the public backlash against LtG originates not from methodological flaws but from ideological resistance. Additionally, subsequent updates employing the recent datasets demonstrate the real-world trajectories track remarkably with LtG scenarios, rebutting many early critiques.

\subsection{Recalibration and Validity}

Since the release of the World3 model, several attempts have been made to reassess its validity and compare its trajectories with empirical data. An empirical validation of the World3-03 model compared its results with real-world data from 1970-2000, focusing on three scenarios: BAU, CT, and SW \cite{turner_comparison_2008}. The analysis demonstrated that observed global trends align closely with the BAU scenario if growth continues unmitigated. However, the effectiveness of technological advancement was questioned in its ability to reverse systemic stress highlighting how solutions such as biofuel, energy efficiency, and others often generate unintended consequences, reinforcing rather than mitigating environmental and economic pressure \cite{turner_comparison_2008}. These warnings that appear timely given the continued pattern of claims that proven negative consequences of data center development will be offset by unproven promises of positive consequences that are yet to materalize \cite{iea_energy_2025}. 

Another recalibration of the World3-03 model was performed with empirical data from 1995-2012, specifically in the context of Scenario 2 from the original study (i.e., BAU2) \cite{pasqualino_understanding_2015}. The paper identified key deviations from Scenario 2: 1) industrial-sector productivity has been lower than anticipated; 2) the service sector has significantly outperformed expectations, accelerated by information technology and a growing finance industry; 3) increased food production is attributed to technological innovation with reduced impact on land erosion. These discrepancies suggest that structural changes such as significant shifts towards services and decoupling food outputs from land use are already alleviating stresses highlighted in the original model.

In 2021, a revisit to the LtG was performed using empirical data up to 2020 with projections from the updated World3-03 model \cite{herrington_update_2021}. The analysis delved into four scenarios as described in Table \ref{tab:tab1}. Contrary to \cite{turner_comparison_2008}, \cite{herrington_update_2021} that argued that the original BAU scenario is closest to the observed trends, it also found that BAU2 and CT emerge as the most closely aligned with empirical data. The CT and BAU2 scenarios differ in how sharply they predict decline: BAU2 presents a severe collapse pattern, while CT exhibits decline under highly optimistic assumptions about the effect of technological advancement. Thus \cite{herrington_update_2021} warns that assumptions under CT might be overly optimistic. The paper also highlights that the SW scenario on the sustainable outlook is the least consistent with empirical data.

The most recent recalibration of the World3-03 model was performed by adjusting 35 variables and optimizing the model against eight key empirical datasets using an iterative algorithm. The findings state that the recalibrated trajectories lie between BAU and BAU2 scenarios \cite{nebel_recalibration_2024}. The new trajectories suggest that collapse still occurs due to resource depletion rather than pollution overshoot, as per BAU2, but occurs fifty years later and reaches a higher peak compared to the earlier BAU. The paper \cite{nebel_recalibration_2024} thus echoed the message from the LtG on the persistent risk of systemic collapse and aligned with findings from \cite{turner_comparison_2008}.

\subsection{Computing within Limits}

The \textit{Computing within Limits} community established the need to shift focus towards the ecological and planetary boundaries rather than scaling performance 10 years ago, and it continues to be a concern as is seen with the current AI trajectory \cite{chen_computing_2015}. Limits to growth as a concept has been used to explore and quantify the ``computational limits of deep learning'' \cite{thompson_computational_2023} and more recently AI scaling trends and the resulting consequences \cite{bhardwaj_limits_2025}. Computing has also been considered as ecocide because of the carbon emissions and energy footprint associated with computation, the combined social and ecological impacts of computing infrastructure that cause resource scarcity, the materiality of computing including severe damages that are possible from improper handling of e-waste, and the tech hype and techno-solutionism that makes computing a facilitator of ecocide \cite{comber_computing_2023}. 

Our paper contributes to the aims of the \textit{Computing within Limits} community to ``[explore] ways that new forms of computing [supports] well-being while enabling human civilizations to live within global ecological and material limits'' by presenting a method with which to quantify the impacts of AI scaling (by exploring data center development) to limits to growth \cite[p.~86]{nardi_computing_2018}. Thus we revisit the limits to growth concept that grounded the conception of \textit{Computing within Limits}.

\section{Methods}

We first explain our rationale for simulating data center development using the World3-03 model and explain its relevance. We then discuss how we updated the World3-03 model including the reasons for our methods, modelling details, and an illustration of the model parameters and relationships. Finally, we outline a few challenges that we encountered in the process of updating the World3-03 model. 

\subsection{Rationale for AI in the World3-03 Model}

It has been found that the ICT sector contributes between 2.1\% and 3.9\% of global carbon emissions, prompting the need to integrate its growing emissions into broader climate models and policymaking \cite{freitag_real_2021}. Similarly, the growing carbon footprint of emerging technologies including AI, blockchain, and IoT is a significant concern as these technologies demand high computational power and infrastructure expansion \cite{freitag_real_2021}. AI systems, and specifically GenAI, demand significant computational power leading to high electricity consumption and increased water usage for cooling \cite{kaack_aligning_2022,Li_Yang_Islam_Ren_2023}. Growing climate impacts have been pinpointed along the entire AI supply chain, from hardware manufacturing to data center operations \cite{luccioni_environmental_2024}. 

These factors can significantly affect the global development trajectories thus prompting an assessment of its potential impact. The World3-03 model is a widely recognized system dynamics model that is ideally suited to capture the reinforcing loops, delays, and potential tipping points that AI growth may trigger. Accordingly, we explore whether explicitly incorporating data center development to the World3-03 model will demonstrate a change in dynamics between causal relationships. 

\subsection{Updates to the World3-03 Model}

There are several methods that can be used to update the World3-03 model: adding new variables, refining technology parameters, updating model assumptions, or reconstructing AI impacts as a separate sector and connecting it with the existing model. 

Prior to explaining our method for updating the World3-03 model, we provide clarifications around the use of key terminology. We define \textbf{variables} broadly as dynamic quantities that change over time, and \textbf{parameters} as constants or fixed values that define the characteristics of the system and influence how variables interact. \textbf{Equation} is defined to be a mathematical expression that defines relationships between different components of the system. \textbf{Sector} is defined as a major subsystem that represents a broad functional area, e.g., non-renewable resources, which contains stocks, flows, feedback loops and variables. 

\begin{figure*}
    
    \begin{subfigure}[b]{0.9\textwidth}
        \includegraphics[width=\textwidth]{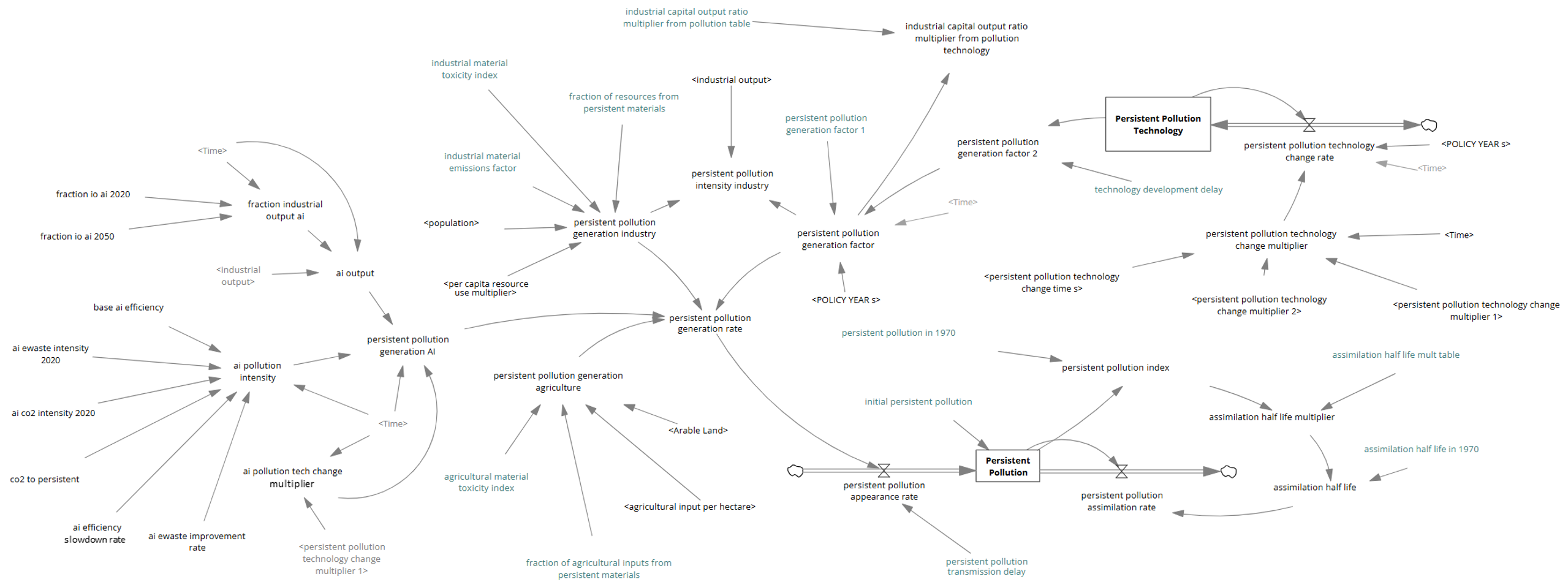}
        \caption{The persistent pollution sector in the World3-03 model with new variables (left cluster) added to simulate pollution caused by AI scaling.}
        \label{fig:fig1a}
    \end{subfigure}
    
    ~ 
    \begin{subfigure}[b]{0.9\textwidth}
        \includegraphics[width=\textwidth]{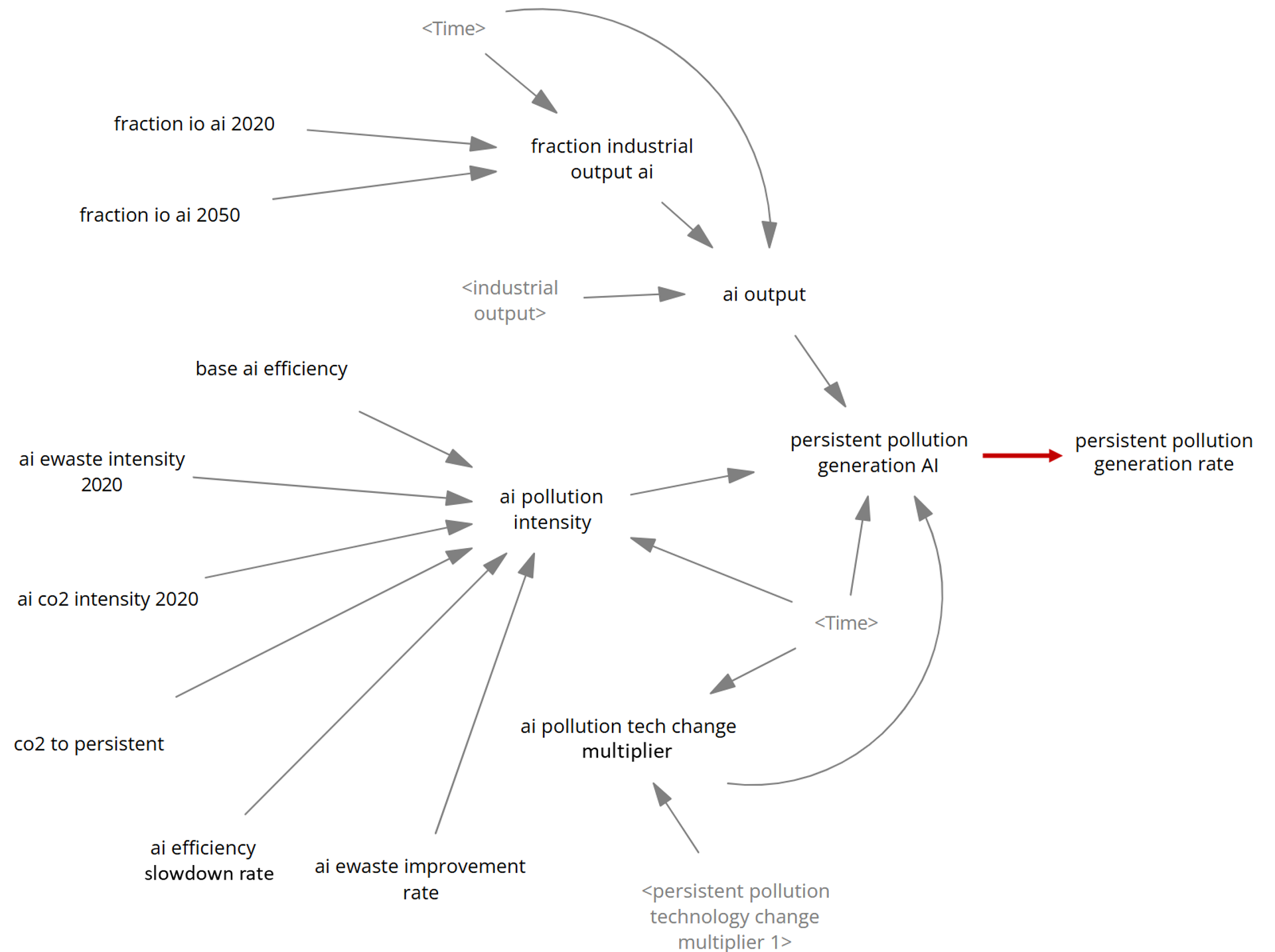}
        \caption{The added variables to simulate pollution caused by AI scaling are connected with the existing variables in the persistent pollution sector through a relationship between variables: ‘persistent pollution generation AI’ and ‘persistent pollution generation rate’.}
        \label{fig:fig1b}
    \end{subfigure}
    \caption{ A screenshot of the Vensim Software \cite{ventana_systems_inc_vensim_2025} showing the persistent pollution sector along with added AI variables in \textbf{(a)} full view and \textbf{(b)} focused on the relationship connecting new variables to existing variables.
    }
    \label{fig:fig1}
    \vspace{-4mm}
\end{figure*}

Our chosen method to explore how rapid scaling of AI, through materials and operations of data centers, impacts limits to growth was to update the model by adding variables into the existing persistent pollution sector using the Vensim PLE 10.3.2 software \cite{ventana_systems_inc_vensim_2025}. We specifically make our updates to the 2004 version of the model (i.e., World3-03) and in the paper, we use the term `World3-03' when discussing the 2004 version and `World3' when discussing previous versions. The new variables include both parameters and equations, as well as new connections that were formed between select existing parameters. The entire persistent pollution sector including our additions is pictured in Figure \ref{fig:fig1a}. Figure \ref{fig:fig1b} shows a closer view of the new variables and how they connect to the existing variables in the persistent pollution sector (through a relationship between variables: `persistent pollution generation AI' and `persistent pollution generation rate'). We list the additions in Table 5 (description of variables, both parameters and equations), Table 6 (the values of the parameters), and Table 7 (equations) in Appendix A. 

We chose to embed AI-relevant variables inside the existing persistent pollution sector of the World3-03 model given the substantive research that currently shows the impact of AI on CO\textsubscript{2} emissions \cite{luccioni_estimating_2023,luccioni_counting_2023}. Furthermore, debates around carbon footprint dominate current regulatory and corporate areas as principal terms of environmental impact \cite{mahapatra_assessment_2021}. Thus updating the persistent pollution sector was possible given the availability of current data (on how AI impacts emissions). Additionally, this method would introduce the least amount of errors as pollution impacts are already considered within the dynamics simulated by the World3-03 model (as compared to more complex methods discussed next). 

In contrast, updating technology-relevant coefficients would be challenging given the difficulty in obtaining how certain values in the model were calculated. Another option would have been to build an AI sector. This is more comprehensive but requires extensive data collection and modeling and thus was considered beyond the scope of this paper. 

The constants for the parameters were identified based on both recent academic and non-academic sources reporting potential, future AI scaling (see discussion on this in Section \ref{takeaways}) \cite{Patterson_2021,Nordic_Computer,OECD_2019,Anand_2017,Jacques_2018,IEEE_2020,Thompson_Spanuth_2021,Forti_2020}. Wherever possible we grounded the parameter constants in peer-reviewed literature with others sourced from industry reports; thus the parameters were back solved from the published aggregates. Accordingly, inferred values should be treated as first-pass approximation and not as definitive benchmarks. 

\subsection{Integrating AI Trends into World3-03} 
The World3-03 model simulates dynamic relationships over a temporal scale and as such computes pollution over time. In order to simulate AI's impacts on pollution, we could not simply create a percentage increase directly but rather had to add dynamic factors that represented AI-caused pollution over time. Adding variables in this manner mirrors the structure of the World3-03 model. 

The persistent pollution sector of the BAU scenario of the World3-03 model takes into account two sources of impact (industrial activity and agriculture) through intensity factors, sums the resulting flows, and lets that total accumulate in a single long-lived stock. In the BAU scenario, persistent pollution generation agriculture is calculated by multiplying the total arable land by the agricultural input per hectare, then scaling by the fraction of agricultural inputs from persistent material, and finally adjusting by the agricultural material toxicity index to account for long-term environmental impact. In parallel, persistent pollution generation industry is calculated by multiplying population by per capita resource use multiplier, then scaling by the fraction of resources from persistent materials, and finally applying both the industrial materials emissions factor and industrial toxicity to determine the overall pollution impact. To represent AI impacts without disturbing the integrity of these original feedback loops, we introduced the pathway for AI that follows the logic of the industrial and agricultural activity. 

Thus following this logic, persistent pollution generation for AI is calculated as a fraction of industrial output spent on AI, scaled carbon, and e-waste footprint. The empirically grounded share of global industrial output is earmarked for the AI industry. The earmarked economic activity is translated into persistent pollution units by combining 1) an initial operational and embodied carbon coefficient, 2) a parallel coefficient reflecting the lifecycle toxicity of electronic waste, and 3) a conversion constant that maps physical emissions onto the model’s dimensionless pollution scale. Both coefficients are allowed to decline at rates calibrated to historical trends in compute efficiency and materials circularity. However, they are bounded below by the same global floor that constrains industrial and agricultural intensities, thereby preserving cross-sectoral comparability. The resulting flow is then added directly to the legacy industrial and agricultural pollution flows before entering the persistent pollution stock. Since all other sectors and parameters remain untouched, any deviation from the canonical trajectories can be attributed unambiguously to the added AI variables. This targeted, transparent augmentation equips World3-03 model to be used to explore whether continued expansion (or policy-driven moderation) of AI development will accelerate, delay, or avert the overshoot dynamics identified in earlier runs, maintaining the model’s original coherence.

We executed two scenarios with the World3-03 model over the interval 1900-2100 with a 0.5-year time step. The first scenario was the standard scenario which is the baseline that corresponds to the BAU and the second was the AI-augmented scenario with our added variables. From each scenario, we extracted diagnostic outputs for persistent pollution, i.e., the stock of long-lived pollutants, and human ecological footprint (HEF), i.e., the ratio of humanity’s resource throughput to the Earth regenerative capacity. We detail added parameters in Table 5, parameter values in Table 6, and equations in Table 7 in Appendix A. 

\subsection{Challenges}

While the initial methods by which the model can be updated appear elementary, there are several nuances involved. We first attempted to update the World3-03 model by adding a new sector on AI development which would begin in 2020. However, considering that building the AI sector in the World3-03 model requires more data inputs and careful consideration of equations, we scaled back to single sector modifications. 

Another challenge was in calibrating the constants for the parameters because empirical estimates about data center operations and materials usage, along with other AI impacts, can vary by an order of magnitude depending on the sources, and particularly in industry reports. Thus we selected mid-range baseline values (detailed in Table 6, Appendix A). Additionally, due to the lack of data, some parameter values have been derived from the existing figures, and should be considered as approximations rather than definitive values. 

We viewed and modified the model through the Vensim PLE software \cite{ventana_systems_inc_vensim_2025} which generates ``warnings'' that do not impede being able to run the model but are nevertheless recommended to resolve. We evaluated and resolved several such warnings in the process of updating the model.  For example, LOOKUP\_BOUNDS warnings occurred because the model requested a value from the lookup table outside the defined bounds of the table. While running the BAU scenario, the warnings were generated for 12 variables. To improve the validity of the model, LOOKUP\_BOUNDS warnings have been treated by increasing the bounds. 

\section{Findings}

We present findings from a comparison between our AI-augmented scenario and the BAU scenario. Based on our added variables to the persistent pollution sector of the World3-03 model, we are able to see updates to the persistent pollution stock. Overall, when compared to the BAU scenario, the AI-augmented scenario results in higher pollution. Comparisons are reported at five benchmark years, i.e., 2020, 2040, 2060, 2080, 2100 and as percentage differences relative to BAU, described further below and listed in full in Table 8 in Appendix A. 

We below show how impacts from AI development can be translated into World3-03 stock and flow language. Our objective was not to deliver a point forecast of a future scenario but rather to visualize the kinds of questions that can be asked from the model. We frame our findings below as a proof of concept rather than a definitive prediction. 

We first demonstrate that even simulating one minor modification, i.e., that AI affects pollution, the system behaviour was updated (i.e., the persistent pollution stock). To summarize, the persistent pollution stock peaked later and was 4\% higher in the AI-augmented scenario as compared to BAU, leading to 45\% larger residue in 2100. As a result of this changed behaviour, further changes were triggered in another variable, human ecological footprint (HEF) where HEF remained approximately 11\% above the BAU baseline. (While many other variables may have been updated as a result of the persistent pollution stock, we report only the changes to HEF given its direct relevance to pollution, discussed further below). These magnitudes are not decisive predictions, but rather they are signposts that the direction of change matters and the feedbacks remain as relevant today as Meadows et al. emphasized in 2004 \cite{meadows_limits_2004}.

\subsection{Variable 1: Persistent Pollution}

Figure \ref{fig:fig2} demonstrates that the persistent pollution stock’s output begins to diverge visibly in 2030 for the BAU and AI-augmented scenarios. While a 4\% higher peak may appear small, recent advances in attribution science clarify the massive implications of each additional percentage point, even when merely considered in monetary terms: ``Each extra percentage point contribution to total 1850–2020 CO\textsubscript{2} and CH\textsubscript{4} emissions generates a further \$834 billion in global economic losses from extreme heat in 1991–2020'' \cite[p.~897]{callahan_carbon_2025}. World3-03’s non-linear feedbacks amplify small stocks which may then magnify into sizable human-welfare impacts. The AI-augmented scenario peaks later and higher, then falls more slowly. In World3-03, the persistent pollution stock feeds into population and food systems through health and land productivity multipliers. A later higher peak is postponed, yet essentially deepens the downstream feedback that triggers the collapse. 



\subsection{Variable 2: Human Ecological Footprint}

We additionally explore the updates to the HEF variable given its direct relevance to pollution and updated output that is triggered from the changes to the persistent pollution stock. HEF is defined as ``the sum of three sectors: the arable land used for crop production in agriculture; the urban land used for urban-industrial-transportation infrastructure; and the amount of absorption land required to neutralize the emission of pollutants, assumed to be proportional to the persistent pollution generation rate'' \cite[p.~293]{meadows_limits_2004}. The comparison between the BAU and AI-augmented scenarios, in Figure \ref{fig:fig3}, shows that because AI-caused pollution and material throughput intensify feedbacks of BAU, the AI-augmented scenario descends a little faster initially, yet it remains consistently 4-8\% above BAU through to 2040. Cumulatively (area between the curves) the world spends approximately 7\% more biocapacity than in the baseline over 2020-2070. The AI scenario finishes the century 11\% above BAU and still marginally in overshoot. The implication is that, even after collapse, the planetary debt left by an AI-heavy economy is larger and persists longer.



\begin{table*}[ht]
\begin{minipage}[b]{0.48\textwidth}
    \centering
    \includegraphics[width=\textwidth]{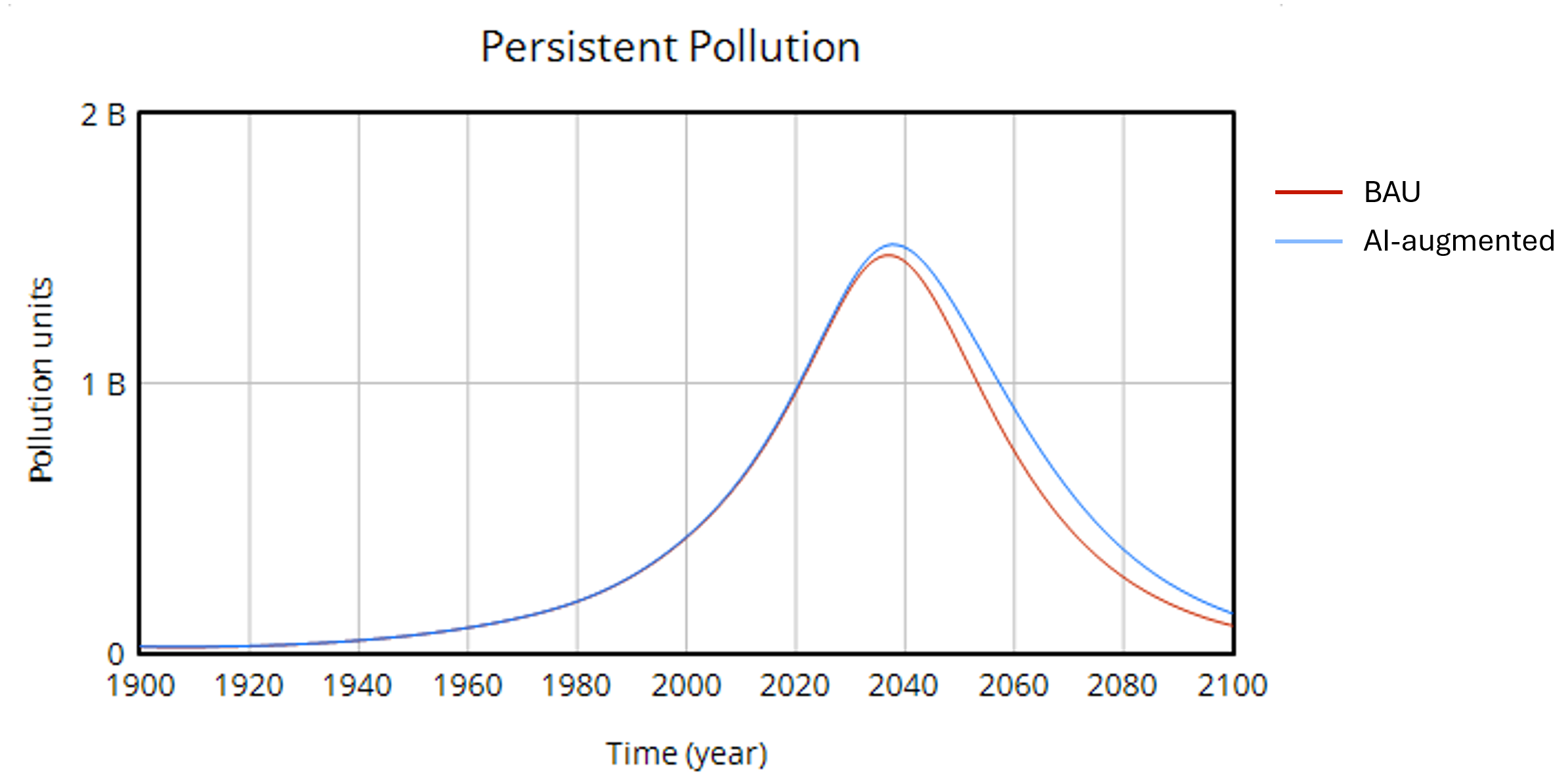}
    \captionof{figure}{Standard and AI-augmented scenarios for Persistent Pollution Stock.}
    \label{fig:fig2}
  \end{minipage}
  \quad
  \begin{minipage}[b]{0.48\textwidth}
    \centering
    \includegraphics[width=\textwidth]{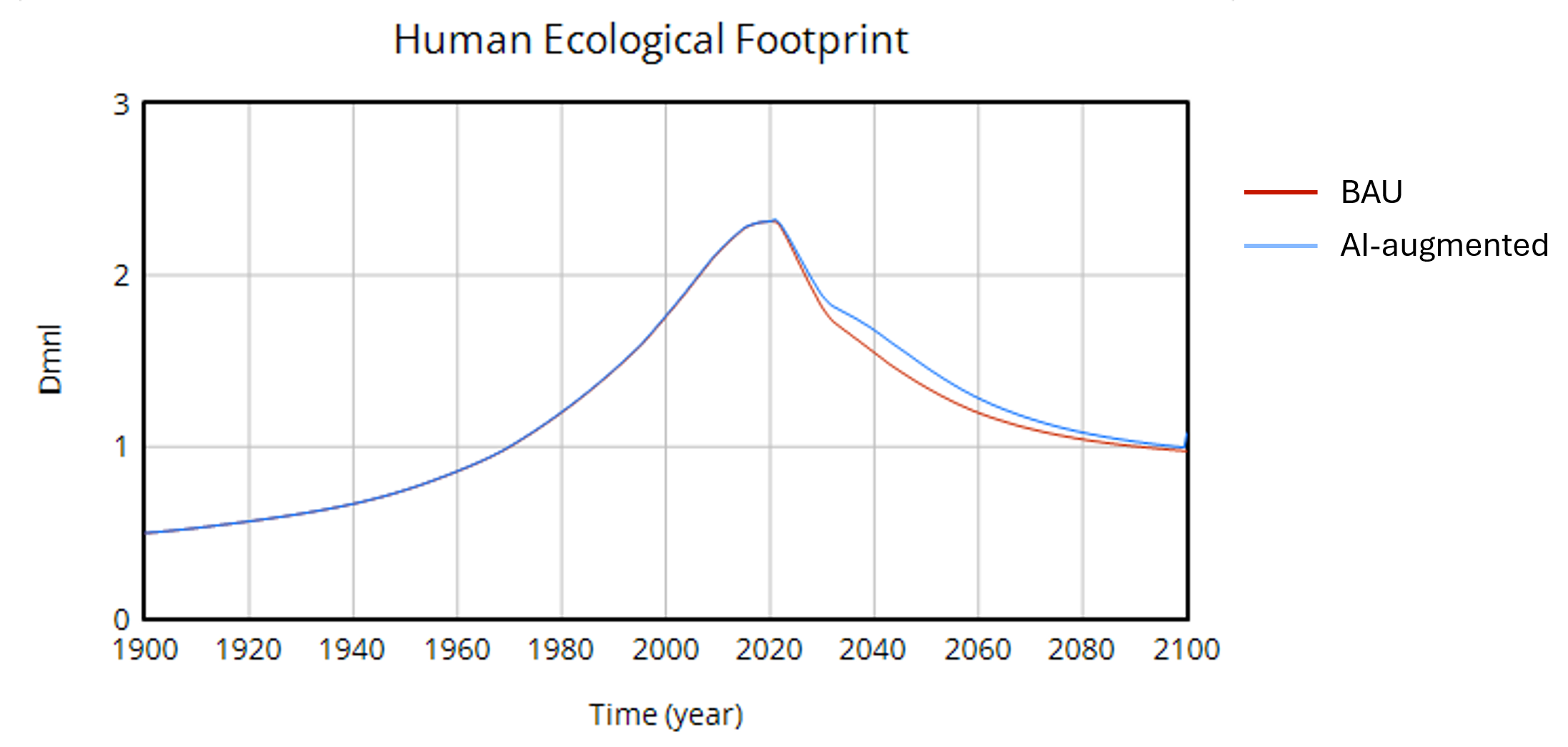}
    \captionof{figure}{Standard (BAU) and AI-augmented scenarios for Human Ecological Footprint}
    \label{fig:fig3}
  \end{minipage}
\end{table*}

\begin{table*}[ht]
\begin{varwidth}[b]{0.4\textwidth}
    \centering
    \begin{tabular}{ l r r r }
      \toprule
      Year & \% Change from BAU \\
      \midrule
      2020 & 0.94  \\
      2040 & 3.77  \\
      2060 & 21.69  \\
      2080 & 37.31 \\
      2100 & 45.35  \\

      \bottomrule
    \end{tabular}
    \caption{Percent change between BAU and AI-augmented scenarios for Persistent Pollution Stock.}
    \label{tab:tab2}
  \end{varwidth}%
  \quad
  \begin{varwidth}[b]{0.4\textwidth}
    \centering
    \begin{tabular}{ l r r r }
      \toprule
      Year & \% Change from BAU \\
      \midrule
      2020 & 0.01\%  \\
      2040 & 8.40\%  \\
      2060 & 7.09\%  \\
      2080 & 3.83\% \\
      2100 & 10.71\%  \\
      
      \bottomrule
    \end{tabular}
    \caption{Percent change between Standard (BAU) and AI-augmented scenarios for Human Ecological Footprint.}
    \label{tab:tab3}
  \end{varwidth}%
\end{table*}

\section{Discussion}

Including AI impacts as modifications in World3-03 opens the door to new questions that the current version cannot answer. In this section, we first discuss the key takeaways from our findings. Second, we discuss the limitations of our study by highlighting that our results demonstrate the \textit{viability} of using the World3-03 model to quantify relationships between AI impacts and world development but do not yet quantify those relationships itself. Lastly, we outline several research opportunities to encourage the use of the World3-03 model to explore relationships between AI and other variables in a quantitative manner. The research opportunities are ways to deepen the understanding of how AI development interacts with planetary limits by modifying the World3-03 model such as by: updating different variables than we have presented, adding new scenarios, adding new subcomponents, and simulating AI development projections.

\subsection{Takeaways} \label{takeaways}

The modifications made to \textit{increase} persistent pollution and the resulting modification to HEF demonstrate that our added assumptions, drawn from evidence, enable us to modify the World3-03 model and obtain expected consequences: AI scaling may deepen and prolong global overshoot in comparison to the Standard Run (BAU). This may further suggest that adding an AI-related layer to a growth-oriented global economy widens the gap between humanity's demands and Earth’s regenerative capacity. Our results based on primitive, preliminary explorations of how AI impacts can be incorporated into the World3-03 model demonstrates that this is a promising method to more rigorously quantify the dynamics of the relationships among various systems. We outline several research opportunities to explore these relationships in a later section.  

The World3-03 model simulates scenarios based on provided data and assumptions. In current discussions around AI’s impact on the environment, there are two asymmetrical approaches that are taken. On the one hand, climate scientists and researchers report on observed and predicted environmental impact of AI using exhaustively researched and empirically validated measures based on conservative assumptions. On the other hand, Big Tech and its supporters make `climate-positive' assurances with no empirical evidence (e.g., as seen in \cite{ketan_joshi_is_2025}) and instead create hype \cite{floridi_why_2024,varoquaux_hype_2025} around future transformative abilities of AI. In the meanwhile, data center operations have precise measurements of their energy and water impact but such information is purposefully obfuscated \cite{Jeans_2021,Gabbott_2024}.

\subsection{Limitations}

As a proof of concept paper, our objective was to show that the World3-03 model can be re-examined and applied to include AI-related variables, thereby projecting global dynamics in a new light. Future researchers should consider several limitations that delineate the scope of our modifications.

We modified the persistent pollution generation rate, while other variables, such as labour productivity, industrial output, land use, and resource depletion, remained untouched. Hypothetically, these could affect the model as strongly as, or even more strongly than, pollution. Therefore, the resulting output is incomplete by design: it illustrates only one aspect of AI’s footprint. These untouched sectors constitute key areas for further exploration, as outlined in the research opportunities section.

AI-related industries are fast moving and constantly evolving. Empirical estimates have widened substantially over the past five years, and consequently, every coefficient could shift by an order of magnitude as new evidence emerges. The values used for the model, as well as the numerical outputs, should be perceived as illustrative ranges, not calibrated predictions. They may appear pessimistic or optimistic depending on which predictions ultimately prove correct.

On the technical side, the software provided two warnings for current agriculture inputs and food ratio. The warnings suggested a mismatch between the initial conditions set manually and the calculated values based on the model equation. Despite several attempts to treat these warnings, a resolution was not found. However, we do not believe this to impact our findings as the differences were minuscule and therefore unlikely to affect the model dynamic over a long-term horizon. 

World3-03 is a globally aggregated model, thus its feedback loops may not capture real-time demands or surface regional pressures on planetary boundaries. While such simplifications keep the equations tractable, they blur regional nuances – precisely the hotspots where limits are likely to be breached first and where targeted policy could be most effective.

\subsection{Research Opportunities}

\noindent\textbf{Modifying different variables:} Further exploration of updating the World3-03 model to simulate AI impacts on limits to growth can include modifying and adding variables other than pollution, such as non-renewable resources and land development. Data center development demands vast quantities of high-grade silicon and rare-earth elements \cite{luccioni_environmental_2024}, thus integrating those flows into the non-renewable-resource stock would reveal potential early-warning breaking points in World3-03. At the same time, the construction of data centers and related logistics infrastructure is converting land at a rapid pace \cite{Gabbott_2024,Hogan_2015}. Embedding a variable representing the impacts of AI development on land use would allow researchers to assess whether terrestrial limits emerge simultaneously with, or even precede, pollution and resource constraints. The updated model could then test scenarios that explore the limits associated with unconstrained AI scaling.

\noindent\textbf{Establishing scenarios for AI scaling:} Understanding how AI might reshape global trends within planetary boundaries demands more than a single calibrated scenario; it requires a structured set of what-if experiments. Given the diverse range of projections of ecological impacts, potential efficiency gains, and more, it may be useful to establish scenarios of AI scaling in which different ranges of values can be considered for parameters. 

We therefore propose a scenario family that can be explored within the current framework: Standard Scenario, Rapid AI Growth, Regulated Growth with a Sustainability Focus, and AI Stagnation. The Standard Scenario (i.e., BAU), unchanged from the classic model, would act as the control against which the remaining three AI scenarios can be compared. The three AI scenarios would be as follows: 
\begin{itemize}
    \item Rapid AI scaling: fast resource extraction and high pollution accompany soaring productivity, leading to intense resource use and a potential overshoot-and-collapse dynamic. This case investigates current acceleration narratives and the trade-off between short-term gains and long-term stability.
    \item Regulated growth with sustainability focus: strict limits on energy use, possibly combined with a shift to renewables, slow deployment but aim for a more balanced, durable trajectory.
    \item AI stagnation: regulatory caps or raw-material shortages curb adoption, offering a lower-bound estimate of resource use and pollution.
    \item Efficiency gains and rebound effects: Most positive consequences of AI fall under efficiency gains in other sectors \cite{iea_energy_2025}. Incorporating these, and their rebound and other effects, would lend much expressive power to the models at the cost of significant complexity.

\end{itemize}

\noindent\textbf{Adding new AI sector:} Another promising research direction is to build a stand-alone AI sector with its own stocks (e.g., AI capital, compute capacity, critical material inventories, etc.) and flows (hardware turnover, e-waste generation, electricity demand, etc.) and then connect its key variables to the existing World3-03 loops. This modular approach would place each AI variable in a single, transparent location, making its influence on the wider system easy to trace. This method would allow researchers to tune the AI sector without disturbing unrelated sectors, turning the model into a living toolkit for exploring how AI scaling may accelerate or alleviate pressure on planetary boundaries.

\noindent\textbf{Simulating current AI projections:} A next stage research agenda could be a scenario that links the AI capital stock to industrial output and, in turn, captures the feedback between investments and economic growth. In essence, this new scenario, with explicit AI capital stock from venture capital and corporate investments, would feed directly into World3-03 loops. This scenario will visualize how current financial projections and investments in AI might push or relax the planet’s biophysical limits. Feeding ``bullish'' capital investment trajectories for the AI share into the model could reveal how rapidly the global system tips into collapse.

\noindent\textbf{Moving towards planetary boundaries and pathways:} The World3-03 model offers none or very limited considerations of geographic specifics, the dynamics of planetary boundaries and tipping points, and other planetary ecosystem dynamics \cite{rockstrom_planetary_2009,rockstrom_planetary_2024,steffen_planetary_2015,williams_boundary_2024}. Future work should explore planetary boundaries and Shared Socioeconomic Pathways (SSPs) as modelling and scenario baseline for the questions explored here, moving beyond limitations of the World3-03 model \cite{riahi_shared_2017,oneill_roads_2017}. 

\section{Conclusion}

AI systems impact all lives on this planet in economic, social, and ecological ways. A system dynamics approach through a planetary-scale World3 model offers a practical method for exploring macro-level consequences. This includes quantitative modelling of impacts on planet-scale limits. 

If AI scaling ever were to reach the absurd `growth' conjured up by Big Tech figures -- as for example Eric Schmidt's recent claim that AI will eventually demand `99 percent' of the world's electricity consumption \cite{Wilkins_2025} -- it would most certainly shatter planetary boundaries. This makes it an urgent challenge to \textit{Computing within Limits}.

In this paper, we have demonstrated how the World3-03 model still remains a feasible method for simulating system-level dynamics of limits to growth, including how the accelerating development and operations of hyperscale data centers at a global scale impacts these limits. Using an exploratory approach, we added variables to the World3-03 model that would increase pollution based on recent evidence and prediction of its acceleration due to AI scaling. Accordingly, we were able to simulate an increase in pollution figures in the World3-03 model and thus established a proof of concept for the World3-03 model’s continuing relevance. Future applications of the model can provide a quantitative approach to exploring ‘what-if’ questions regarding future AI and computing impacts. 
 
Currently, predictions and metrics surrounding AI often boast of its innovative abilities while lacking critical distance and empirical evidence, and industry players obscure the details surrounding its harms. Here we take the opposite approach: we begin by modelling what is widely documented, i.e., the lifecycle impacts. Once beneficial impacts can be proven, they should also be integrated.


\bibliographystyle{ACM-Reference-Format}
\bibliography{LIMITS25}



\section*{Supplementary material}

# Appendix

Table 4: Scenarios adapted from the LtG of 1972 [10].

| Scenario | Name | Description |
|---|---|---|
| 1 | Standard Run | The "standard" world model assumes no major shifts in the systems shaping global development. Food, industry, and population grow fast until a declining resource base halts industrial growth. Delays cause pollution and population to rise briefly after the peak. Growth stops as death rates climb due to less food and healthcare. |
| 2 | Natural Resource Reserved Doubled | To test the model's resource assumption, 1900 reserves were doubled, with all other conditions unchanged from the Standard Run. Industrialization reaches a higher peak as resources last longer, but the larger industrial base generates pollution faster than the environment can absorb. Pollution spikes, driving up death rates and cutting food outputs. By the end, resources are nearly exhausted, despite the initial doubling. |
| 3 | "Unlimited" Resources | Resource depletion is avoided by assuming unlimited nuclear energy and double resource reserves, enabling widespread recycling and substitution. But if these are the only changes, growth still ends due to increasing pollution. |
| 4 | "Unlimited" Resources and Pollution Controls | A new technological change was added to reduce pollution per unit of output to one-fourth of its 1970 level, aiming to solve earlier resource and pollution issues. This allows growth until arable land becomes the limit. Food per person drops, and industry slows as capital shifts to agriculture. |
| 5 | "Unlimited" Resources, Pollution Controls, and Increased Agricultural Productivity | To prevent the earlier food crisis, land yield was doubled in 1975, alongside prior pollution and resource policies. These combined changes remove key growth limits, leading to high population and industrial levels. But despite lower pollution per unit, total output grows so much that pollution still triggers collapse. |

| 6 | "Unlimited" Resources, Pollution Controls, and "Perfect" Birth Control | Instead of boosting food output, the model tests more effective voluntary birth control to ease the food issue. Population still grows, just more slowly, and the crisis is only delayed by 10–20 years. |
|---|---|---|
| 7 | "Unlimited" Resources, Pollution Controls, Increased Agricultural Productivity, and "Perfect" Birth Control | Four tech policies are added to avoid collapse: full resource use with 75% recycling, pollution cut to one-fourth, land yield doubled, and widespread access to birth control. Still, industrial growth stops and death rates rise as resources run low, pollution builds, and food drops. |
| 8 | Stabilized Population | This scenario matches the standard model, except population is stabilized after 1975 by balancing birth and death rates. The unchecked industrial feedback loop drives continued exponential growth in output, food, and services. Eventually, resource depletion causes industrial collapse. |
| 9 | Stabilized Population and Capital | Capital growth is limited by making investment equal to depreciation, added to the population stabilization policy. This halts exponential growth and creates a temporary steady state. But without resource-saving technology, high levels of population and capital still drain resources quickly. As resources fall, industrial output drops, and capital becomes less efficient, with more used for extraction than production. |
| 10 | Stabilized I | Technological and growth-limiting policies are combined to create a long-term sustainable equilibrium. Tech measures include recycling, pollution control, longer capital life, and soil restoration. Value shifts emphasize food and services over industry. Births equal deaths, and industrial investment matches depreciation. Industrial output per person reaches three times the 1970 average. |
| 11 | Stabilized II | If strict growth limits are lifted and population and capital are regulated within system delays, equilibrium shifts to higher population and lower industrial output per person. With perfect birth control and a desired family size of two by 1975, the birth rate slowly nears the death rate due to age-structure delays. |

| | | | |
|---|---|---|---|
| 12 | Stabilizing Policies Introduced in the Year 2000 | | When 1975 policies are delayed until 2000, the equilibrium becomes unsustainable. Population and industry grow too high, leading to food and resource shortages before 2100. |

Table 5: New and existing variables for the persistent pollution sector of World3-03.

| Variables | Type | Description | Importance |
|---|---|---|---|
| *co2_to_persist* | New Parameter | An empirical conversion constant used in World3 to translate a mass of greenhouse gases into the model's abstract persistent-pollution units (pu). | It is essential because without the conversion all new AI terms would carry incompatible units. |
| *ai_co2_intensity_2020* | New Parameter | Direct operational, plus embodied $CO_2$ released per dollar of AI infrastructure and model-training expenditure in 2020. | Captures the energy/$CO_2$ footprint of hyperscale data-centres and other build-outs. This variable is a key leverage point for climate-aware computers. |
| *ai_ewaste_intensity_2020* | New Parameter | Life-cycle $CO_2$-equivalent to toxic metals, Per-and polyfluoroalkyl substances (PFAS), and others in AI hardware discarded in 2020. | This variable has been added because AI's footprint is not only energy but also material toxicity. |
| *frac_io_ai_2020, frac_io_ai_2050* | New Parameter | Fraction of industrial output of AI in 2020 and 2050. These variables are the empirically anchored share of global industrial output devoted to AI hardware & services in 2020 and its saturation level in 2050. | These levers, not intensity, drive how big the AI sector becomes inside the macroeconomy. |
| *industrial output* | Existing variable | This stock is the global dollar flow generated by the industrial capital sector. | When a recession, capital shortage, or overshoot-and-collapse episode drives industrial output down, the amount of compute that can be bought also falls. Thus, AI- related pollution rises and falls consistently with |

| | | | the rest of the economy. |
|---|---|---|---|
| *base_ai_efficiency* | New Parameter | This variable is an annual fractional improvement in compute efficiency, i.e. how quickly AI hardware is getting more energy-efficient over time. | This is a negative exponent, which means that as AI gets more efficient, it uses less electricity for the same economic output which reduces $CO_2$ emissions per dollar of AI services. |
| *ai_efficiency_slodown_rate* | New Parameter | Decay of the improvement rate. | Without a slowdown, AI could appear to become infinitely clean, which is unrealistic. |
| *ai_ewaste_improvement_rate* | New Parameter | This parameter lowers the AI_Ewaste_intensity part of AI_pollution_intensity each year to reflect progress in circular design, reuse and recycling. | It is essential because, even if computers get cleaner, material toxicity can dominate the long-lived pollution stock. |
| *Fraction_industrial_output_ai* | New Equation | Time-path of the AI share. | This variable helps avoid a discontinuous jump in 2020 and keeps the capital-allocation loop numerically stable. |
| *ai_output* | New Equation | This variable converts a share of the global economy into an absolute activity level. | It links the AI branch to the existing capital feedback: if the world economy collapses, AI activity collapses with it. |
| *ai_pollution_intensity* | New Equation | This variable scales today's footprint to any future year before the tech multiplier is applied. | It allows experimenters to tune learning-by-doing, i.e. make the 3 % and 5 % slopes steeper or shallower without touching the conversion logic. |
| *persistent_pollution_technology_change_multiplier* | Existing variable | Global S-curve lookup that halves effective intensity every ~25 yr down to a floor of 0.5. | Re-used so AI inherits the same "planetary clean-tech progress" assumption. |

| *ai_pollution_tech_change_multiplier* | New Equation | This variable helps prevent AI from falling below the global floor. | This lets the model decide whether AI is held to the same floor or allowed to fall lower/higher in scenario studies. |
| --- | --- | --- | --- |
| *persistent_pollution_generation_ai* | New Equation | The new inflow that feeds the long-lived pollution stock. | It activates exactly when generative-AI deployment begins. Automatically fades after 2100, so late-century behaviour is not distorted. |
| *Time* | Existing variable | This variable reflects the year. | It helps to set the time bounds within other variables. |

Table 6: Values for the New Parameters
The intention of providing the parameters and their values is not to assert their universal validity but rather to make the assumptions behind them explicit. By articulating these assumptions transparently, the table below allows us to evaluate, revise, and update the parameters in order to develop alternative versions.

| Parameters | Suggested Value | Units | Definition | Justifications and Assumptions |
|---|---|---|---|---|
| co2_to_persist | 2.3 e-4 | Dmnl | Fraction of a one-off $CO_2$ pulse that remains in the atmosphere/ocean "long enough" to show up in the LtG persistent pollution stock. | Multi-model reviews state that 20-35% of the $CO_2$ remains in the atmosphere for centuries after equilibration with the ocean [2]. Taking a midpoint 0.23 of the values above provides a "persistent" fraction. The World3 model treats persistent pollution in abstract "pollution units" [11], so any real-world emission flow should be converted to the scale via calibration factor to keep the magnitudes comparable with the model's existing industrial and agricultural inflows. Thus, normalization constantly has been applied to keep the legacy loop consistent. This parameter is not a pure physical constant, but a hybrid conversion factor that preserves compatibility with the World3 model. Key assumptions for this parameter are: using a fixed midpoint fraction; assuming persistence is unchanging over time and emission scale; treating different pollutant classes as equivalent once expressed in persistent units. |

| ai_CO2_intensity_2020 | 1.5 e-1 | Dmnl | Direct operational $CO_2$ emitted per 2020-$ of AI output. | The AI-specific load is around 30-50 TWh [9], which is used as a proxy for early-2020s intensity. Multiplying this by the global 2020 average power-grid intensity of about 0.45 $tCO_2$/MWh [4], then divided by AI market revenues of around US $156.5 billion [12] in 2020, yields emission intensity of 0.086-0.144 kg $CO_2$ per $, which rounded up to 0.15 for modeling. The literature indicates that emissions from AI workloads can vary by orders of magnitude depending on data center efficiency, processor type, and grid mix, which can shift footprints by 100-1000x or 5-10x [16]. This variability highlights that the calculated value is an approximation rather than precise constant. Key assumptions include: 2023 data center TWh loads are representative of 2020 AI workloads; the global average grid intensity reflects the actual hosting mix; dividing by market revenues is a valid scaling even though many revenues are "non-compute"; embodied emissions from chip manufacturing and hardware are excluded. This methodology assumes that AI market revenues serve as a reasonable proxy for electricity-consuming AI activities, though this introduces uncertainty since US $156.5 billion includes other revenue streams with potentially minimal direct electricity correlation. The limitations are that AI market revenues are broader and load data for 2020 may overstate direct energy use. |

| Parameter | Value | Units | Description | Notes |
|---|---|---|---|---|
| *ai_ewaste_intensity_2020* | 3.5 e-4 | Dmnl | Embodied-carbon, end-of-life & recycling impacts per 2020-$ of AI hardware spend. | The approximation is made by scaling from global data center embodied emissions. The International Energy Agency estimates that data centers and networks together emitted around 330 Mt $CO_2$e in 2020 [19]. The embodied energy, i.e. the total energy consumed during the entire lifecycle, represents approximately 18% of total data [17] center energy consumption. To focus only on hardware, the assumption has been made that 6-10% of total emissions are hardware embodied, i.e. ~20-30 Mt $CO_2$e per year. The global data-center IT capex is ~US $200-250 billion [13] in 2020, of which 60—80B(~15-25%) is attributable to AI hardware. Dividing ~25Mt $CO_2$e embodied by US $70 billion AI spend yields approximately 3.5 e-4. The figures used in the calculation of this parameter are approximations that add compounding uncertainty, e.g. actual embodied share of 6-10% can vary widely based on the stakeholder. Additionally, AI-specific hardware capex shares are also uncertain and vary year by year.<br>The figures for the parameter are limited by the information about uncertain global AI spend and corporate reporting boundaries. This aligns with wider literature that suggests that embodied hardware emissions are often underestimated and may be higher than commonly assumed [14]. The parameter should be interpreted as order of magnitude, system-level intensity rather than a precise value. |

| Parameter | Value | Units | Description | Notes |
|---|---|---|---|---|
| *frac_io_ai_2020* | 1.3 e-2 | Dmnl | Share of total industrial output that is AI/compute in 2020. | This parameter is calculated by dividing the global AI market revenue of US $156.5 billion [12] in 2020, by total manufacturing value added of US $13.6 trillion [20]. The assumptions for this parameter are: AI market revenue is assumed proportional to industrial output; price basis consistency assumes that both revenue and manufacturing data are reported in comparable USD without distortions from inflation. The limitations are that it is hard to separate data center revenue streams from AI-specific infrastructure. Measuring the size of the AI economy is challenging because the revenues overlap with broader ICT services, cloud platforms, and analytics markets, which reinforces the uncertainty using market revenue as proxy for industry output [15]. |
| *frac_io_ai_2050* | 6.0 e-2 | Dmnl | Target share of industrial output devoted to AI by 2050. | The parameter is calculated by dividing projected AI market revenues by total global industrial output. Using OECD projection of US $311 trillion [8] for world GDP in 2050, multiplied with World Bank's industry and construction value-added share of approximately 26% of GDP [20], yields a global industrial output of around US$ 80 trillion by 2050. The projection of a US $5 trillion AI market by 2050 [21], when divided by US $80 trillion produces the target parameter value. The independent industry studies suggest that AI could contribute much more to the global economy and GDP growth annually [1, 7]. This highlights the large uncertainty surrounding this projection. The figures used for the calculation present uncertainty. The calculation is based on several assumptions: industry remains 26% of GDP through 2050; US $5 trillion forecast is representative. |

| Parameter | Value | Units | Description | Notes |
|---|---|---|---|---|
| *base_ai_efficiency_improvement* | 2.5 e-1 | Dmnl | Annual fractional improvement in compute efficiency. | AI hardware efficiency is rising by roughly one order of magnitude(~10x) from the mid-2010s to about 2020 for training and inference devices. A tenfold gain over a decade implies a compound annual improvement of 10^1/10 – 1, which is roughly round down to 0.25 [6]. This baseline is consistent with the post-Dennard CPU trend that implies the improvement at roughly constant power by 2x every 3.5 years, i.e. 2^1/3.5 -1 = 0.21 [5]. The calculation assumes that efficiency improves smoothly and uniformly across all AI workloads, even though real-world gains occur in discontinuous steps when new architectures are released. Additionally, the parameter reflects only operational energy efficiency, not total system-level efficiency. |
| *ai_efficiency_slowdown_rate* | 4.0 e-2 | 1/year | How fast the above improvement rate decays. | The annual improvement rate of chip performance-per-dollar fell from 48%/yr (2000-2004) to about 8%/yr(2008-2013. Interpreted linearly over around a decade, that decline of (0.48 – 0.08)/10, which is roughly equal to 0.04 [18]. The parameter assumes that the calculation assumes the slowdown proceeds linearly, even though the progress is uneven, with innovation booms and periods of decline. It also applies figures derived from general-purpose computing directly to AI accelerators, which may have different scaling dynamics. |

| Variable | | | | | |
|---|---|---|---|---|---|
| ai_ewaste_improvement_rate | 3.0 e-2 | 1/year | Annual reduction in e-waste CO$_2$-eq per $ due to circularity & reuse.. | Hyperscale AI data centers are already reusing a large share of components by keeping servers in service longer, extending lifetimes from 4 to 6 years, reusing and refurbishing more parts, using recycling materials. It is projected that the global e-waste will rise from 53.6Mt in 2019 to 74Mt by 2030 [3]. This increase corresponds to a compound annual growth rate of (((74/53.6)^1/11-1), which equals 3% per year. This parameter reflects the approximate annual efficiency gain needed to match the project growth rate of global e-waste, preventing persistent pollution pressure beyond current trends. The limitations of these values should be also considered: AI-relevant data centers may follow faster turnover cycles, global recycling rates of AI hyperscale remain low. | |

Table 7: New and updated equations.

| Variable | Equations | Unit |
|---|---|---|
| fraction_industrial _output_AI | = fraction io ai 2020 <br> + (fraction io ai 2050 - fraction io ai 2020) / (1 + EXP(-(Time - 2035) / 5)) | Dmnl |
| ai_output | = IF THEN ELSE(Time>= 2020, fraction industrial output ai * industrial output, 0) | $/year |
| ai_pollution_intensity | = ( ai co2 intensity 2020 <br>     * (1 + EXP(-base ai efficiency * (1 - MAX(0, Time - 2020) * ai efficiency slodown rate))) <br>     * MAX(0, Time - 2020) <br>  + <br>     IF THEN ELSE( <br>     Time < 2070, <br>     ai ewaste intensity 2020 * (1 - ai ewaste improvement rate * MAX(0, Time - 2020)), <br>     2e-05 <br>     ) <br> ) <br> * co2 to persistent | Pollution units/$ |

| | | |
|---|---|---|
| *ai_pollution_tech_ change_multiplier* | = MIN( 5 ,     MAX( 0.7       * persistent pollution  technology change multiplier 1,  0.3    * (1 + 0.1 * (Time- 2020)) ) ) | Dmnl |
| *persistent_pollution_ generation_ai* | = IF THEN ELSE( Time< 2020 :OR:      Time  > 2100 ,    0 ,    ai output     * ai pollution intensity      / ai pollution tech change        multipliyer ) | Pollution units/year |
| *persistent pollution generation rate (existing)* | = ( persistent pollution generation industry      + persistent pollution generation agriculture + persistent pollution generation AI) * ( persistent pollution generation factor ) | Pollution units/year |

Table 8: Comparison between the standard scenario (BAU) and AI-augmented scenario.

| Time (year) | Persistent Pollution: AI-Augmented | Persistent Pollution: BAU | % Change from BAU |
|---|---|---|---|
| 2020 | 976172000 | 967120000 | 0.94 |
| 2025 | 1.176E+09 | 1165000000 | 0.99 |
| 2030 | 1.369E+09 | 1352590000 | 1.21 |
| 2035 | 1.495E+09 | 1464870000 | 2.03 |
| 2040 | 1.506E+09 | 1451000000 | 3.77 |
| 2045 | 1.413E+09 | 1323730000 | 6.77 |
| 2050 | 1.26E+09 | 1134610000 | 11.08 |
| 2055 | 1.085E+09 | 932903000 | 16.29 |
| 2060 | 910573000 | 748269000 | 21.69 |
| 2065 | 749359000 | 591696000 | 26.65 |
| 2070 | 607107000 | 463889000 | 30.87 |
| 2075 | 486134000 | 361763000 | 34.38 |
| 2080 | 386071000 | 281165000 | 37.31 |
| 2085 | 304886000 | 218014000 | 39.85 |
| 2090 | 239870000 | 168755000 | 42.14 |
| 2095 | 188261000 | 130473000 | 44.29 |
| 2100 | 147537000 | 101504000 | 45.35 |



\end{document}